\documentclass[referee]{aa} 
\def\note #1]{{\bf #1]}}

\begin{document}
  \title{Radial velocity-acceleration curve analysis of}
  \subtitle{the spectroscopic binary stars by the nonlinear regression}

  \author{K. Karami
          \inst{1,3,4}
          \and
          H. Teimoorinia
          \inst{1,2}
           }
   \offprints{K. Karami}

   \institute{Institute for Advanced Studies in Basic Sciences, Gava Zang, P.O. Box 45195-1159, Zanjan, Iran\\
             \and
             Young Research Club of Azad University of Zanjan, P.O. Box 49195-467, Zanjan,
             Iran\\
             \and
             Institute for Studies in Theoretical Physics and Mathematics (IPM), P.O. Box 19395-5531, Tehran,
             Iran\\
             \and
             Department of Physics, University of Kurdistan, Pasdaran St.,
             P.O. Box 66177-15175, Sanandaj, Iran\\
             email: Karami@iasbs.ac.ir; Teimoorinia@iasbs.ac.ir\\
              }

   \date{Received / Accepted}

   \abstract{We will introduce a new method to derive the orbital elements of
   the spectroscopic binary stars. Fitting the adopted radial velocity curves on the measured
   experimental data, the corresponding radial acceleration data
   are obtained. The orbital parameters are estimated by
   the nonlinear least squares of the radial velocity-acceleration
   curves. This method is implemented for the three double line spectroscopic
   binary systems RZ CAS, CC Cas and V1130 Tau. Numerical calculations show that the result are in good agreement
   with the those obtained by others using the method of Lehmann-Filh\'{e}s.

      \keywords{stars: binaries: eclipsing --stars: binaries: spectroscopic
               }
   }
   \maketitle
\section{Introduction}
Determining the orbital elements of the binary stars help us to
obtain the necessary information such as the mass and the radius
of stars which play the important roles during the evolutions of
the stellar structures. Analyzing both the light and the radial
velocity curves deducing from the photometric and the
spectroscopic observations, respectively, yields to derive the
orbital parameters. One of the usual methods to analyze the
velocity curve is the method of Lehmann-Filh\'{e}s (Smart 1990).
Here we introduce a new method to derive these parameters by the
nonlinear regression of the radial velocity-acceleration curves.
We test our method for the three double-line spectroscopic binary
systems RZ Cas, CC Cas and V1130 Tau which have the following
properties.

The Algol system RZ Cas is the semi-detached eclipsing binary
stars with a period $P=1.195$ days. There is a mass transfer
between the two components. Some evidences existed for periodicity
in the period changes due to apsidal motion. RZ Cas is also a
radio source and an X-ray source. See Riazi et al. (1994) and
Maxted et al. (1994) and references therein. CC Cas is the massive
O-type binary system with a period $P=3.366$ days. This system is
a detached binary with the two components within the main sequence
band. It is an actively interacting system. The system shows a
mass loss rate $3\times 10^{-7}$m$_{\odot}$yr$^{-1}$ due to the
stellar winds. Such mass loss from the system yields to a change
in the orbital period by the rate of less than $3\times 10^{-6}$
days per century (Hill et al. 1994). V1130 Tau is a very closed
detached system including the two components with similar size and
brightness. It consists of two almost identical F3 V stars on a
tight orbit with a short period $P=0.799$ days. (Rucinski et al.
2003).

In Sect. 2, we reduce the problem to solving a equation for the
radial velocity-acceleration relation which is a nonlinear
function in terms of the orbital parameters. In Sect. 3 the
nonlinear regression technic for estimating the orbital elements
is discussed. In Sect.4 the numerical results implemented for the
three different binary systems are reported. Section 5 is devoted
to concluding remarks.
\section{Formulation of the problem}
The radial velocity of star in a binary system is defined as
follows
\begin{equation}
V_r=V_{cm}+\dot{Z},\label{Vr}
\end{equation}
where $V_{cm}$ is the radial velocity of the center of mass of
system with respect to the sun and
\begin{equation}
\dot{Z}=K[\cos(\theta+\omega)+e\cos\omega],\label{Zdot}
\end{equation}
is the radial velocity of star with reference to the center of
mass of the binary (see Smart 1990). Note that the dot denotes the
time derivative. In Eq. (\ref{Zdot}), $\theta$, $\omega$ and $e$
are the angular polar coordinate, the longitude of the perihelion
and the eccentricity, respectively. Also
\begin{equation}
K=\Large \frac{2\pi}{P}\frac{a\sin i}{\sqrt{1-e^2}},\label{K}
\end{equation}
where $P$ is the period of motion and the inclination $i$ is the
angle between the line of sight and the normal of the orbital
plane.

The radial acceleration of star can be obtained as
\begin{equation}
\ddot{Z}=-K\sin(\theta+\omega)\dot{\theta}.\label{Zddot1}
\end{equation}
Using Kepler's second law and the relations obtaining for the
orbital parameters in the inverse-square field as
\begin{equation}
\dot{\theta}=h/r^2,\label{thetadot}
\end{equation}
\begin{equation}
h=\frac{2\pi}{P}a^2\sqrt{1-e^2},\label{h}
\end{equation}
\begin{equation}
r=\frac{a(1-e^2)}{1+e\cos\theta},\label{r}
\end{equation}
the radial acceleration, Eq. (\ref{Zddot1}), yields to
\begin{equation}
\ddot{Z}=\frac{-2\pi
K}{P(1-e^2)^{3/2}}\sin(\theta+\omega)(1+e\cos\theta)^2,\label{Zddot}
\end{equation}
where $r$, $a$ and $h$ are the radial polar coordinate, the semi
major axis of the orbit and the angular momentum per unit of mass,
respectively.

Substituting for $\theta$ from Eq. (\ref{Zdot}) in Eq.
(\ref{Zddot}), the result reduces to
\begin{eqnarray}
P\ddot{Z}&=&\frac{-2\pi
K}{(1-e^2)^{3/2}}\sin\big(\cos^{-1}(\dot{Z}/K-e\cos\omega)\big)
\nonumber
\\&&\times\Big\{1+e\cos\big(-\omega+\cos^{-1}(\dot{Z}/K-e\cos\omega)\big)\Big\}^2.
\label{pz:}
\end{eqnarray}
To simplify the notation further, we let $Y=P\ddot{Z}$ and
$X=\dot{Z}$. Hence Eq. (\ref{pz:}) describes a nonlinear relation,
$Y=Y(X,K,e,\omega)$, in terms of the parameters $K$, $e$ and
$\omega$.

One can show that the adopted spectroscopic elements are related
to the orbital parameters. First according to the definition of
the center of mass, the mass ratio in the system is obtained as
\begin{equation}
\frac{m_p}{m_s}=\frac{a_s\sin i}{a_p\sin i},\label{mratio}
\end{equation}
where subscripts $p$ and $s$ return to the primary and the
secondary components of the system.

Then from Kepler's third law and Eq. (\ref{mratio}), the result
reduces to
\begin{equation}
m_p\sin^3i=a_s\sin i(\frac{a_p\sin i+a_s\sin i}{P}
)^2,\label{msin3i}
\end{equation}
where $a$, $P$ and $m$ are expressed in AU, years and solar mass,
respectively. The similar relation is obtained for the secondary
component only by replacing $p$ to $s$ and vise versa, in Eq.
(\ref{msin3i}). Note that in Eqs. (\ref{mratio}) and
(\ref{msin3i}) the parameter $a\sin i$ is related to the orbital
parameters by the aid of Eq. (\ref{K}).
\section{Nonlinear regression of the radial velocity-acceleration curve}
To obtain the orbital parameters $K$, $e$ and $\omega$ in Eq.
(\ref{pz:}), we use the nonlinear regression method. In this
approach, the sum of squares of errors ($SSE$) for the number of
$N$ measured data is calculated as
\begin{equation}
SSE=\sum_{i=1}^{N}(Y_i-\hat{Y}_i)^2=\sum_{i=1}^{N}[Y_i-Y(X_i,K,e,\omega)]^2,\label{SSE}
\end{equation}
where $Y_i$ and $\hat{Y}_i$ are the real and the predicted values,
respectively. To obtain the model parameters, the $SSE$ should be
minimized in terms of $K$, $e$ and $\omega$ as
\begin{equation}
\frac{\partial SSE}{\partial K}=\frac{\partial SSE}{\partial
e}=\frac{\partial SSE}{\partial \omega}=0.\label{SSE-diff}
\end{equation}
To solve Eq. (\ref{SSE-diff}), we use the SAS (Statistical
Analysis System) software. Note that the nonlinear models are more
difficult to specify and estimate than linear models. For
instance, in contrast to the linear regression, the nonlinear
models is very sensitive to the initial guesses for the
parameters. Because in practice, $SSE$ may has to be minimized in
several points in the three dimensional parametric space including
$K$, $e$ and $\omega$. However the final goal is finding the
absolute minimum. Hence choosing the relevant initial parameters
yields to the absolute minimum of $SSE$ which is also stationary.
That means if one change the initial guesses slightly, then the
result reduces to the previous values for the parameters. But if
the regression converged at the local minimum, the model would not
be stationary. See Sen $\&$ Srivastava (1990) and Christensen
(1996). To avoid the problem, two things can help us, first by
taking into account the bound intervals for the parameters which
causes the model to converge quickly. For instance one can put
($0\leq e\leq 1$) for the binary stars which have closed orbits.
The last one is checking the differences between the observational
data and the fitted diagrams using Eq. (\ref{pz:}).
\section{Numerical results}
\begin{figure}
\includegraphics{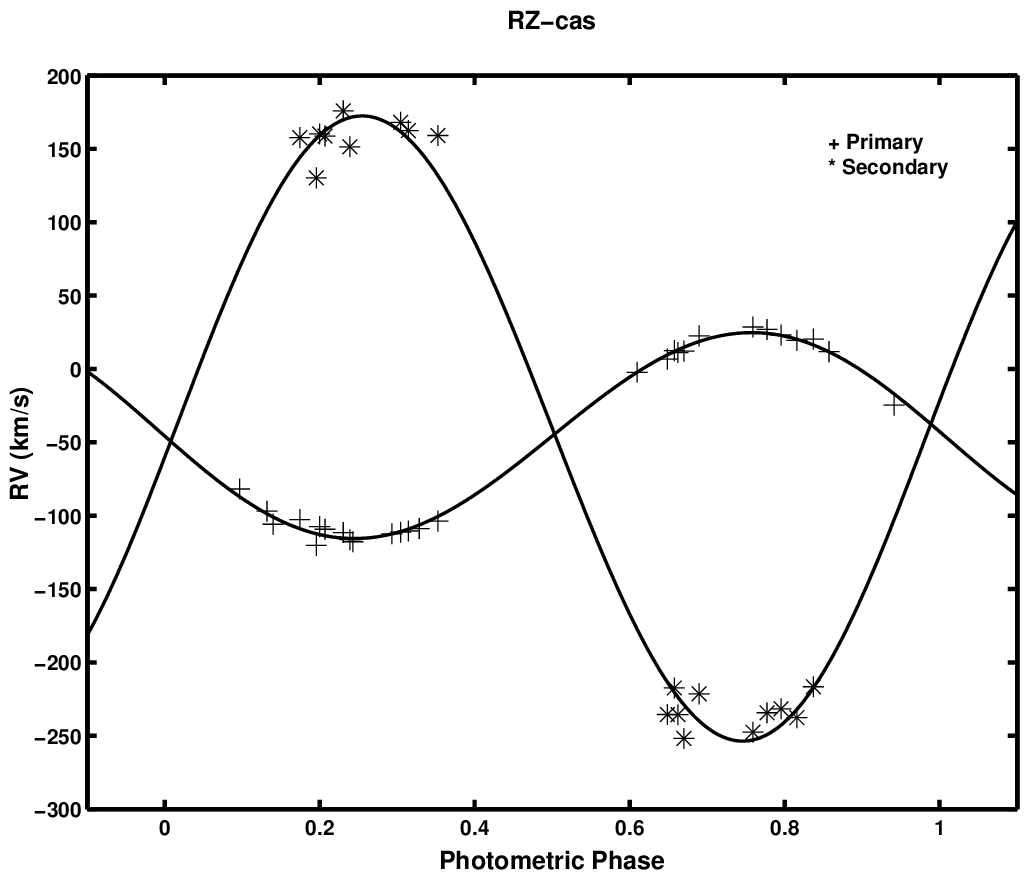}
      \vspace{16cm}
      \caption[]{The radial velocities of the primary and secondary components of
      RZ Cas plotted against photometric phase. The observational data have been measured by Maxted et al. (1994). Legend is given in the right hand up caption.}
         \label{RZ-Cas-RV}
   \end{figure}

\begin{figure}
\includegraphics{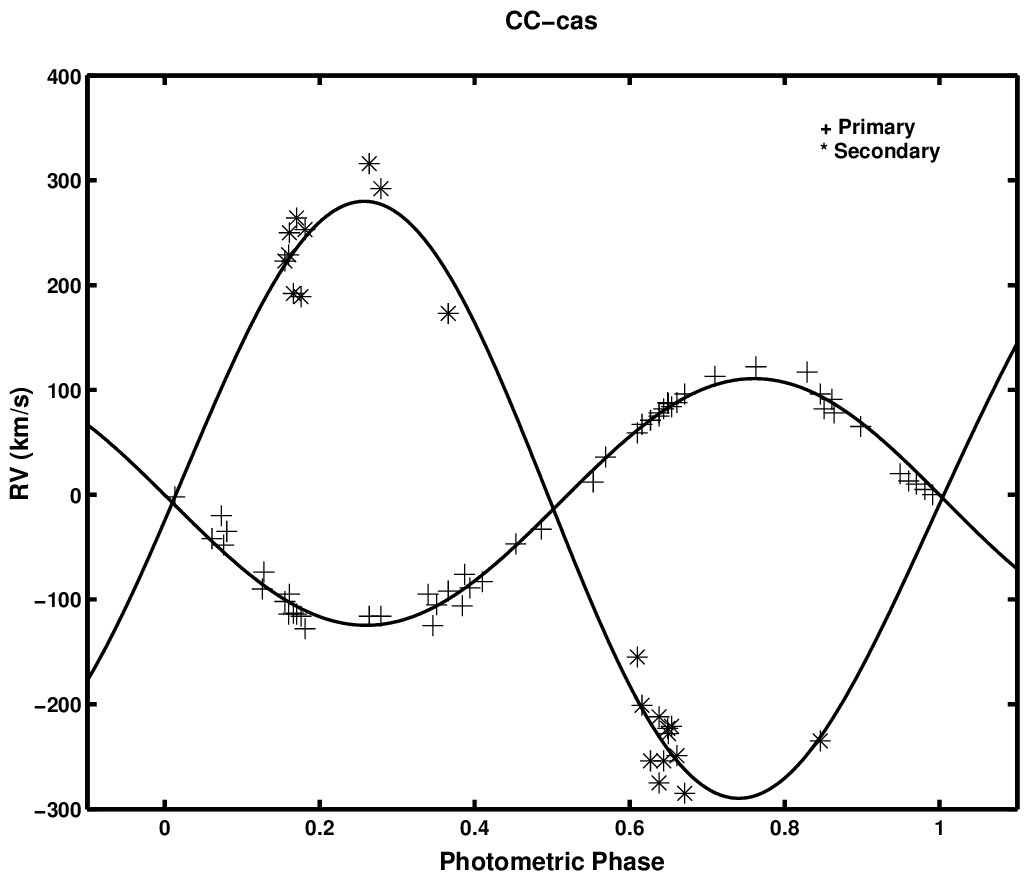}
      \vspace{16cm}
      \caption[]{Same as Fig. \ref{RZ-Cas-RV}, for CC Cas. The observational data are belong to Hill et al. (1994).}
         \label{CC-Cas-RV}
   \end{figure}

\begin{figure}
\includegraphics{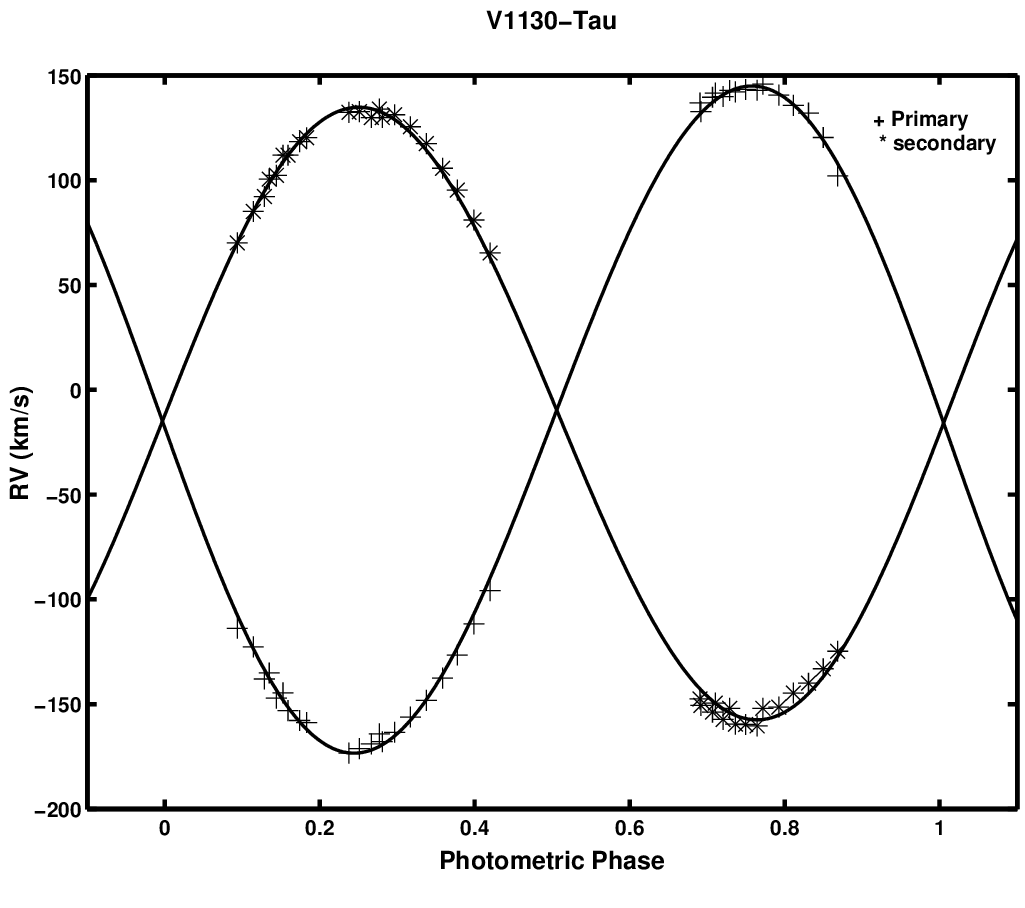}
      \vspace{16cm}
      \caption[]{Same as Fig. \ref{RZ-Cas-RV}, for V1130 Tau. The observational data have been deduced from Rucinski et al. (2003).}
         \label{V1130-Tau-RV}
   \end{figure}

Here we test our new method to derive the orbital and the combined
elements of the binary stars. To do this we consider the three
different double line spectroscopic systems, RZ cas, CC Cas and
V1130 Tau. Using the measured experimental data for the radial
velocities of two components of these systems obtained by Maxted
et al. (1994) for RZ Cas, Hill et al. (1994) for CC Cas and
Rucinski et al. (2003) for V1130 Tau, the suitable fitted velocity
curves are plotted in terms of the photometric phase in Figs.
\ref{RZ-Cas-RV}, \ref{CC-Cas-RV} and \ref{V1130-Tau-RV}. The best
fits are obtained by the polynomials of the intermediate degrees,
e.g., sixth and seventh degrees. Because the low degrees do not
suitably cover the data and high degrees take the fluctuations of
the data which yields to the divergency in the nonlinear
regression of Eq. (\ref{pz:}). The velocity of the center of mass
of the stars are obtained by calculating the areas up and down of
the radial velocity curves. Everywhere these areas become equal to
each other then the velocity of the center of mass is obtained.

\begin{figure}
 \includegraphics{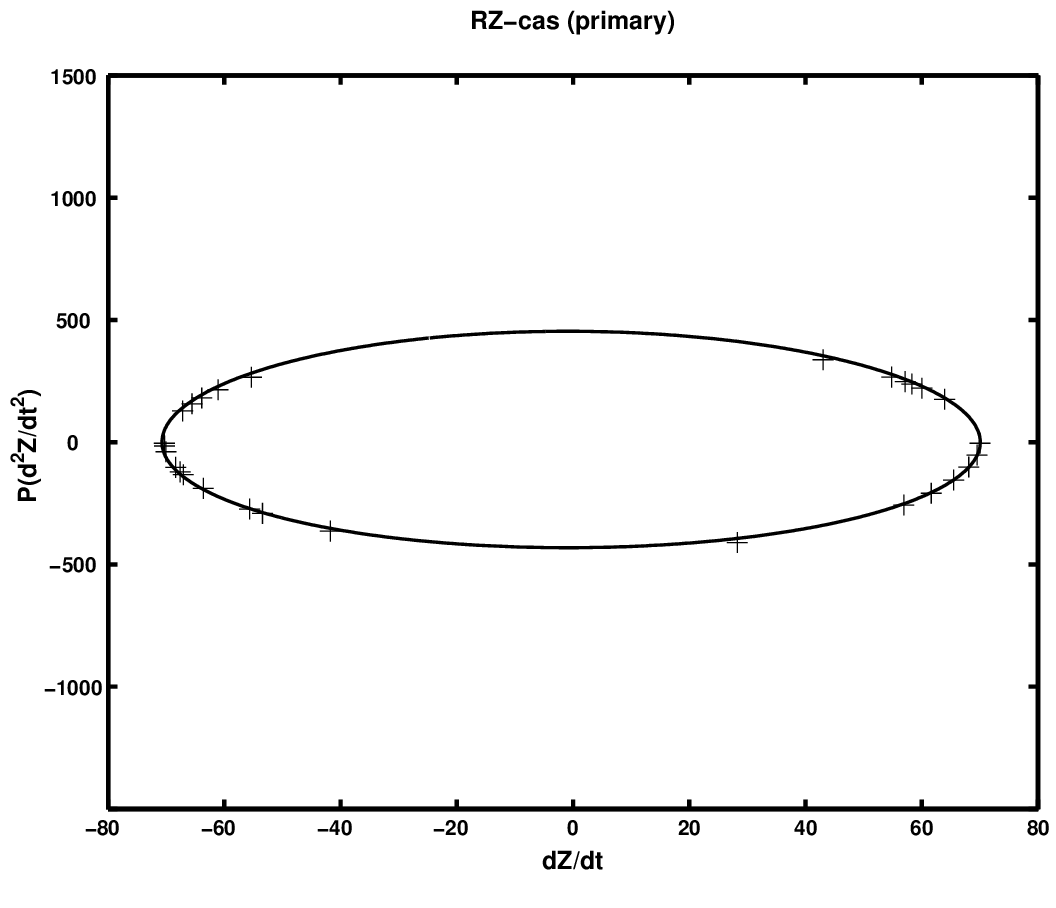}
      \vspace{16cm}
      \caption[]{The radial acceleration scaled by the period versus the radial velocity of the primary
      component of RZ Cas. The solid curve is obtained from the nonlinear regression of Eq. (\ref{pz:}).
      The plus points are
      the experimental data.}
         \label{RZ-Cas-Pri}
   \end{figure}

\begin{figure}
\includegraphics{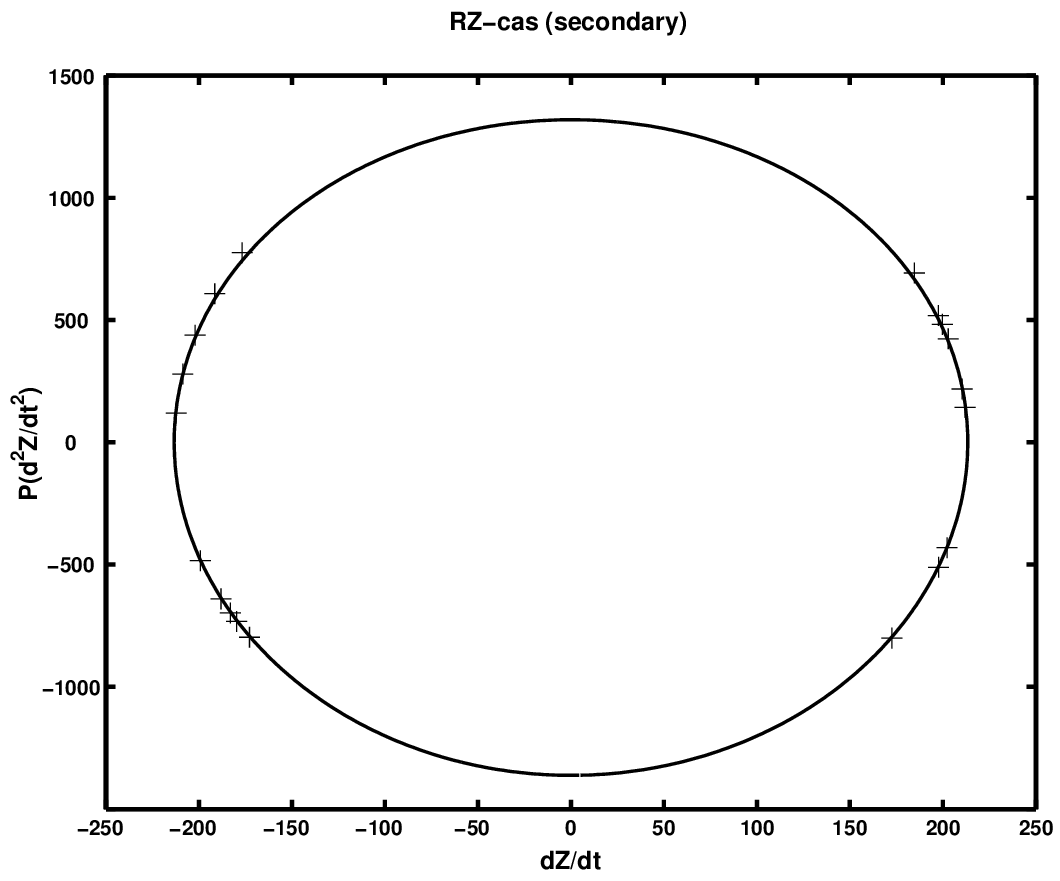}
      \vspace{16cm}
      \caption[]{Same as Fig. \ref{RZ-Cas-Pri}, for the secondray component of RZ Cas.}
         \label{RZ-Cas-Sec}
   \end{figure}

\begin{figure}
 \includegraphics{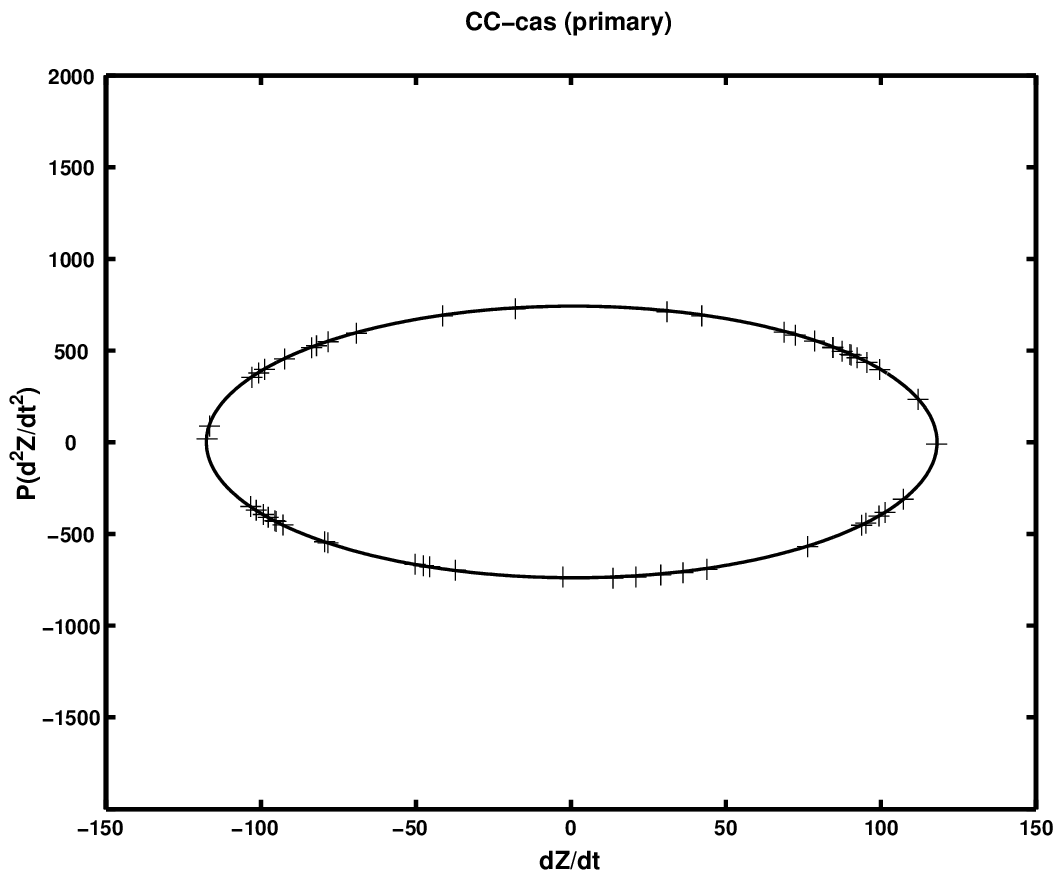}
      \vspace{16cm}
      \caption[]{Same as Fig. \ref{RZ-Cas-Pri}, for the primary component of CC Cas.}
         \label{CC-Cas-Pri}
   \end{figure}

\begin{figure}
\includegraphics{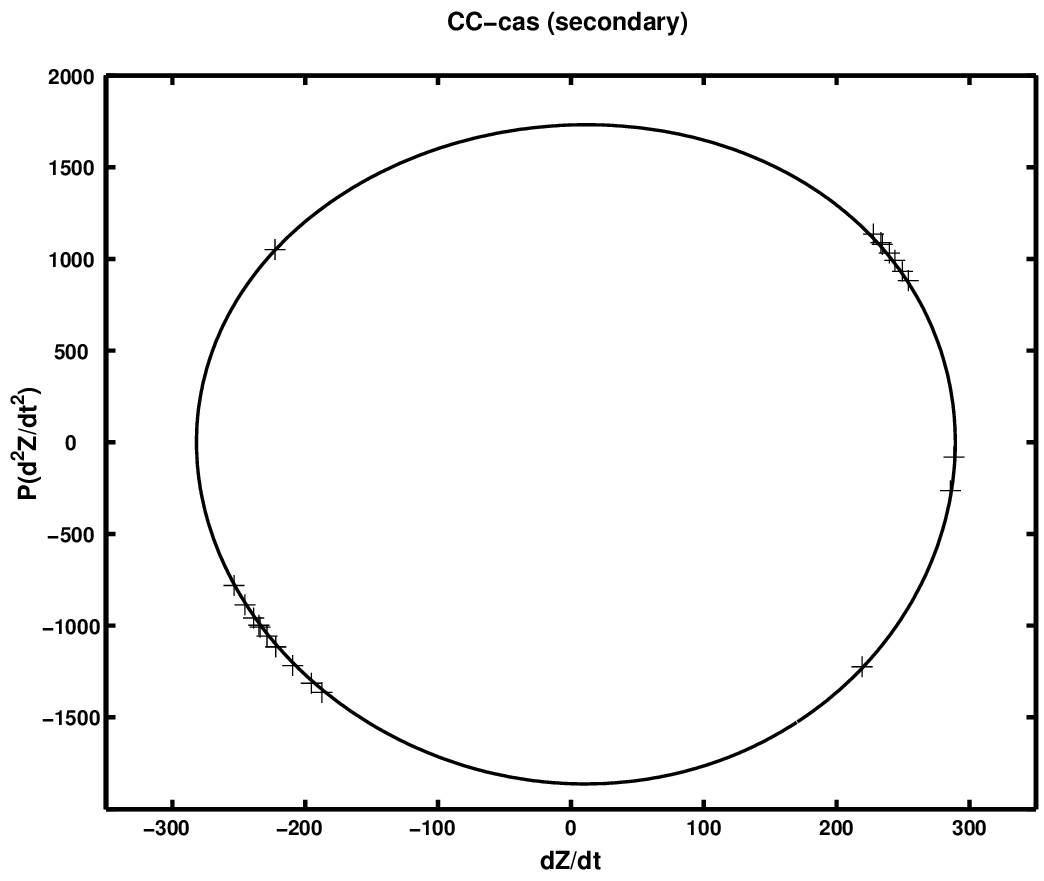}
      \vspace{16cm}
      \caption[]{Same as Fig. \ref{RZ-Cas-Pri}, for the secondary component of CC Cas.}
         \label{CC-Cas-Sec}
   \end{figure}

\begin{figure}
 \includegraphics{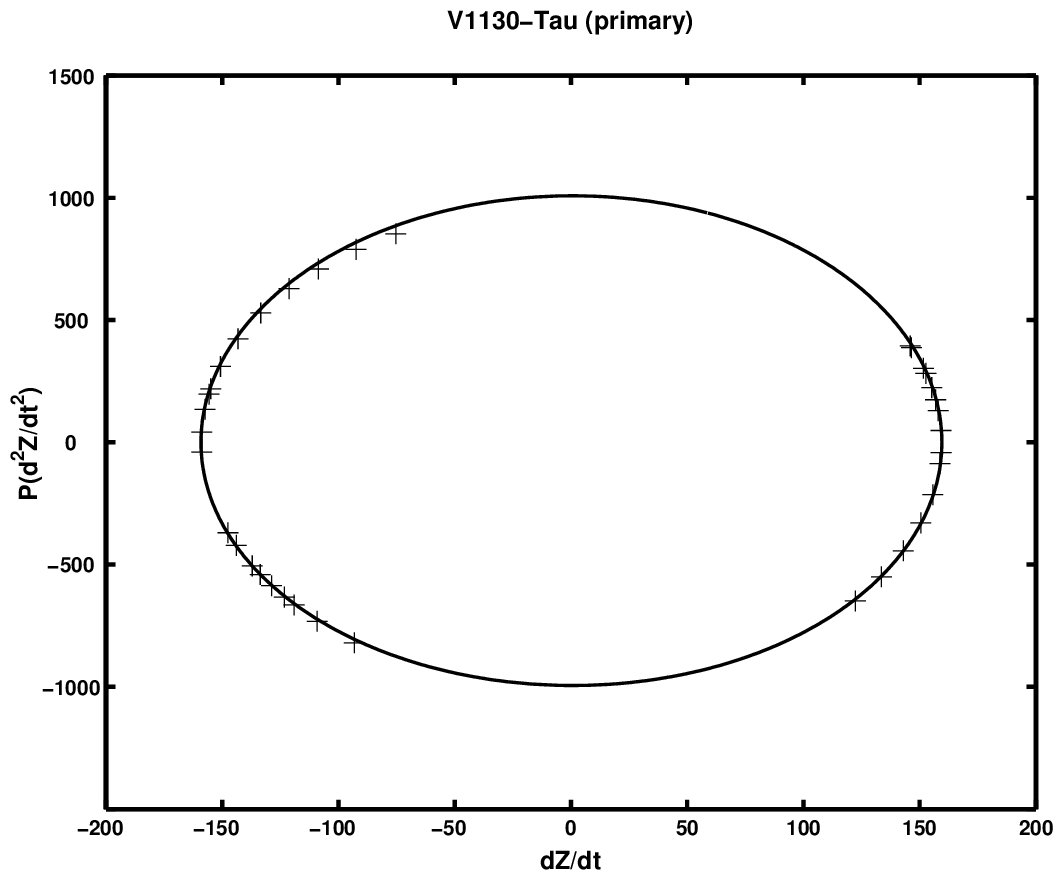}
      \vspace{16cm}
      \caption[]{Same as Fig. \ref{RZ-Cas-Pri}, for the primary component of V1130 Tau.}
         \label{V1130-Tau-Pri}
   \end{figure}

\begin{figure}
\includegraphics{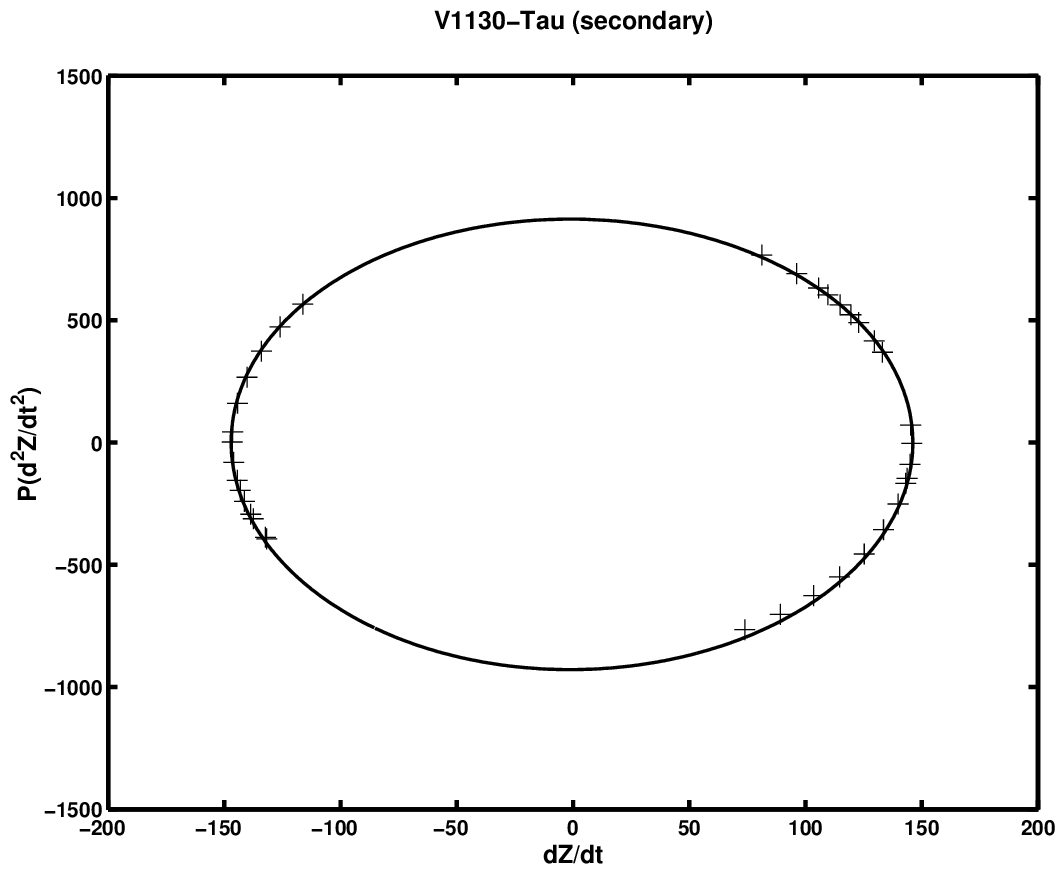}
      \vspace{16cm}
      \caption[]{Same as Fig. \ref{RZ-Cas-Pri}, for the secondary component of V1130 Tau.}
         \label{V1130-Tau-Sec}
   \end{figure}

The radial acceleration data corresponding with the radial
velocity data are obtained by taking the derivative of the adopted
radial velocity curves. Figures $\ref{RZ-Cas-Pri}$,
$\ref{RZ-Cas-Sec}$, $\ref{CC-Cas-Pri}$, $\ref{CC-Cas-Sec}$,
$\ref{V1130-Tau-Pri}$ and $\ref{V1130-Tau-Sec}$ show the radial
acceleration scaled by the period versus the radial velocity data
for the primary and the secondary components of RZ cas, CC Cas and
V1130 Tau, respectively. The solid closed curves are the result of
the nonlinear regression of Eq. (\ref{pz:}), which their good
coincidence with the observational data yields to derive the
optimized parameters $K$, $e$ and $\omega$.

The orbital parameters obtaining from nonlinear least squares of
Eq. (\ref{SSE}) for RZ Cas, CC Cas and V1130 Tau are tabulated in
Tables \ref{RZ-Cas-Orbit}, \ref{CC-Cas-Orbit},
\ref{V1130-Tau-Orbit}, respectively. Table \ref{RZ-Cas-Orbit}
shows that for RZ Cas: 1) $V_{cm}$ and $K_p$ for the primary
component are in good agreement with the those obtained by Maxted
et al. (1994) and Duerbeck $\&$ H\"{a}nel (1979). 2) $e$ and
$\omega$ are closed to the results obtained by Horak (1952). 3)
$V_{cm}$ and $K_s$ for the secondary component are coincide with
the results derived by Maxted et al. (1994). Whereas for $e$ and
$\omega$, we had no any references for comparing.

Table \ref{CC-Cas-Orbit} shows that for CC Cas: 1) $V_{cm}$ and
$K$ for the primary component are in good analogy with the results
obtained by Hill et al. (1994). For the secondary component, $K$
is in relatively well coincidence with the those derived by Pearce
(1927) and Hill et al. (1994). But for $V_{cm}$ is not, however
its interval is compatible with the result of Hill et al. (1994).
The difference may be caused from the existence of high scattering
for the data around the maximum and minimum of the radial velocity
curve (see again Fig. \ref{CC-Cas-RV}).

Table \ref{V1130-Tau-Orbit} shows that for V1130 Tau, 1) for both
components, $K$ is in very good concord with the result estimated
by Ruciniski et al. (2003). 2) For $V_{cm}$, its average value for
the two components is in good agreement with the results of
Rucinski et al. (2003).

The combined spectroscopic elements including $m_p\sin^3i$,
$m_s\sin^3i$, $(a_p+a_s)\sin i$ and $m_p/m_s$ are calculated by
substituting the estimated parameters $e$, $K$ and $\omega$ in
Eqs. (\ref{K}), (\ref{mratio}) and (\ref{msin3i}). The results
obtained for the same three previous systems are tabulated in
Tables \ref{RZ-Cas-Combined}, \ref{CC-Cas-Combined},
\ref{V1130-Tau-Combined}. Tables show that our results are in good
agreement with the those obtained by Maxted et al. (1994), Hill et
al. (1994) and Rucinski et al. (2003) for RZ Cas, CC Cas and V1130
Tau, respectively.

\section{Concluding remarks}
A new method to derive the orbital elements of the spectroscopic
binary stars is introduced. These parameters are obtained from the
nonlinear regression of the radial velocity-acceleration curve of
the system. Using the measured experimental data for radial
velocities of RZ Cas, CC Cas and V1130 Tau obtained by Maxted et
al. (1994), Hill et al. (1994) and Rucinski et al. (2003), we find
the orbital elements of these systems by the mentioned method. Our
numerical results show that the results obtained for the orbital
elements and the combined spectroscopic parameters are in good
agreement with the those obtained by others via the method of
Lehmann-Filh\'{e}s. In a subsequent paper we intend to test
numerically our method for the other different systems.
\begin{acknowledgements}
This work was supported by: Young Research Club of Azad University
of Zanjan, Iran; the Institute for Advanced Studied in Basic
Sciences (IASBS), Zanjan, Iran; the Institute for Studies in
Theoretical Physics and Mathematics (IPM), Tehran, Iran;
Department of Physics, University of Kurdistan, Sanandaj, Iran.
The authors wish to thank Prof. Rucinski, Prof. Hill and Prof.
Maxted for providing valuable consultations.
\end{acknowledgements}

\clearpage
\clearpage
\clearpage
\clearpage
\clearpage
\clearpage
\clearpage
\clearpage
\clearpage
\clearpage
\begin{table}
\caption[]{Spectroscopic orbit of RZ Cas. Values bold type are
assumed. }
\begin{tabular}{lccccc} \hline
 &This Paper  & Maxted et al. (1994) & Hork (1952) &Duerbeck \& H$\ddot{a}$nel (1979)\\\hline
 {\bf Primary} &   \\
$V_{cm} \left(kms^{-1}\right)$ & $-45.08\pm 0.93$ & $-45.5\pm0.7$ &$40\pm2$ & $-45.5\pm1.1$ \\
 $K_p\left(kms^{-1}\right)$&  $70.4\pm0.34$ & $70.9\pm0.8$ &$68\pm2$ & $70.5\pm1.4$ \\
 $e$& $0.013\pm 0.002$ & {\bf 0.000}&$0.01\pm0.03$& $0.024\pm0.023$ \\
$\omega(^\circ)$& $250\pm 11.5$ &--- &$240\pm160$ & $87\pm2$\\
{\bf Secondary} & \\
$V_{cm} \left(kms^{-1}\right)$ & $-40.68\pm 3.69$ & $-41\pm4$ & --- & --- \\
 $K_s\left(kms^{-1}\right)$&  $213.35\pm0.45$ &$213\pm4$ & --- & --- \\
 $e$& $0.008\pm 0.001$ & {\bf 0.000}& --- & --- \\
 $\omega(^\circ)$&$88.84\pm7$&---& --- & ---
\\\hline
\end{tabular}\\
\label{RZ-Cas-Orbit}
\end{table}

\begin{table} \caption[]{Combined spectroscopic orbit of RZ Cas.}
\begin{tabular}{lcc} \hline
 Parameter  & This paper & Maxted et al. (1994) \\\hline
 $m_p\sin^3i/M_\odot$ & $2.126\pm 0.002$ & $2.16\pm0.07$\\
$m_s\sin^3i/M_\odot$ & $0.701\pm 0.005$ & $0.715\pm0.02$\\
$\left(a_p+a_s\right)\sin i/10^6km$ & $4.663\pm 0.007$ & $4.68\pm0.07$\\
$m_p/m_s$ & $3.03\pm 0.02$ & $3.02\pm0.07$\\\hline
\end{tabular}\\
\label{RZ-Cas-Combined}
\end{table}
\begin{table}
\caption[]{Same as Table \ref{RZ-Cas-Orbit}, for CC Cas. }
\begin{tabular}{lccc} \hline
 &This Paper  & Hill et al. (1994) & Pearce (1927) \\\hline
{\bf Primary} &   \\
$V_{cm}\left(kms^{-1}\right)$ & $-7.23\pm2.68$ & $-7.3\pm4.2$ &$-9.9\pm2.4$ \\
 $K_p\left(kms^{-1}\right)$&  $117.87\pm0.51$ & $116\pm1.8$ &$138\pm4$ \\
 $e$& $0.0025\pm0.0005$ & --- &---\\
$\omega(^\circ)$& $330\pm 2$ &--- &--- \\
{\bf Secondary} & \\
$V_{cm}\left(kms^{-1}\right)$ & $-8.63\pm7.31$ & $-12.3\pm6$ & $3.2\pm7$ \\
 $K_s\left(kms^{-1}\right)$&  $285.63\pm1.14$ &$279.4\pm2.3$ & $288\pm9$ \\
 $e$& $0.023\pm0.005$ & ---& --- \\
 $\omega(^\circ)$&$54\pm13$&---&
\\\hline
\end{tabular}\\
\label{CC-Cas-Orbit}
\end{table}

\begin{table} \caption[]{Same as Table \ref{RZ-Cas-Combined}, for CC Cas.}
\begin{tabular}{lccc} \hline
 Parameter  & This paper & Hill et al. (1994) & Pearce (1927) \\\hline
 $m_p\sin^3i/M_\odot$ & $16.2\pm0.2$ & $15.3\pm0.3$ & $18.3\pm1.4$\\
$m_s\sin^3i/M_\odot$ & $6.69\pm0.08$ & $6.3\pm0.2$ & $8.8\pm0.6$\\
$\left(a_p+a_s\right)\sin i/R_\odot$ & $26.83\pm0.08$ & $26.2\pm0.2$ & $28.3\pm0.6$\\
$m_p/m_s$ & $2.42\pm0.02$ & $2.41\pm0.04$ &$2.09\pm0.09$\\\hline
\end{tabular}\\                                                                                                                                                                                                                                                                                                                                                                                                                                                                                                                                                                                                                                                                                                                                                                                                                                                                                                                                                                                                                                                                                                                                                                                                                                                                                                                                                                                                                                                                                                                                                                                                                                                                                                                                                                                                                                                                                                                                                                                                                                                                                                                                                                                                                                                                                                                                                                                                                                                                                                                                      
\label{CC-Cas-Combined}
\end{table}
\clearpage
\begin{table}
\caption[]{Same as Table \ref{RZ-Cas-Orbit}, for V1130 Tau. Number
in parenthesis are errors in the final digits.}
\begin{tabular}{lcc} \hline
 &This Paper  & Rucinski et al. (2003) \\\hline
{\bf Primary} &   \\
$V_{cm}\left(kms^{-1}\right)$ & $-14.45\pm0.89$ & $-12.74(0.46$)  \\
 $K_p\left(kms^{-1}\right)$&  $146.61\pm1.01$ & $147.21(0.63)$  \\
 $e$& $0.005\pm0.003$ & --- \\
$\omega(^\circ)$& $126\pm42$ &--- \\
{\bf Secondary} & \\
$V_{cm}\left(kms^{-1}\right)$ & $-10.9\pm0.9$ & $-12.74(0.46)$   \\
 $K_s\left(kms^{-1}\right)$&  $159.36\pm 0.61$ &$160.11(0.74)$   \\
 $e$& $0.004\pm0.004$ &--- \\
 $\omega(^\circ)$&$292\pm7$&--- \\\hline
\end{tabular}\\
\label{V1130-Tau-Orbit}
\end{table}

\begin{table} \caption[]{Same as Table \ref{RZ-Cas-Combined}, for V1130 Tau.}
\begin{tabular}{lcc} \hline
 Parameter  & This paper & Rucinski et al. (2003)  \\\hline
 $m_p\sin^3i/M_\odot$ & $1.23\pm0.02$ & --- \\
$m_s\sin^3i/M_\odot$ & $1.13\pm0.2$ & --- \\
$\left(a_p+a_s\right)\sin i/R_\odot$ & $4.82\pm0.02$ & --- \\
$m_p/m_s$ & $0.919\pm0.009$ & $0.919(7)$ \\
$\left(m_p+m_s\right)\sin^3i/M_\odot$ & $2.37\pm0.02$& 2.$408(32)$
\\\hline
\end{tabular}\\
\label{V1130-Tau-Combined}
\end{table}
\clearpage
\end{document}